\documentclass[pre,superscriptaddress,tightenlines,twocolumn,showpacs]{revtex4-1}

\usepackage{latexsym,amsmath,amsfonts,tabularx,amssymb,bm}
\usepackage{graphicx,epstopdf}
\usepackage{xspace}

\usepackage{color}

\newcommand{\beginsupplement}{%
        \setcounter{table}{0}
        \renewcommand{\thetable}{S\arabic{table}}%
        \setcounter{figure}{0}
        \renewcommand{\thefigure}{S\arabic{figure}}%
     }

\usepackage{etoolbox}
\patchcmd{\section}
  {\centering}
  {\raggedright}
  {}
  {}
\patchcmd{\subsection}
  {\centering}
  {\raggedright}
  {}
  {}
  
\usepackage{afterpage}

\begin{document}

\title{Long-range nematic order and anomalous fluctuations \\in suspensions of swimming filamentous bacteria}

\author{Daiki Nishiguchi}
\email{nishiguchi@daisy.phys.s.u-tokyo.ac.jp}
\affiliation{Department of Physics, The University of Tokyo, Hongo 7-3-1, Tokyo 113-0033, Japan}

\author{Ken H. Nagai}
\affiliation{School of Materials Science, Japan Advanced Institute of Science and Technology, Ishikawa 923-1292, Japan}

\author{Hugues Chat\'{e}}
\affiliation{Service de Physique de l'Etat Condens\'e, CEA, CNRS, Universit\'e Paris-Saclay, CEA-Saclay, 91191 Gif-sur-Yvette, France}
\affiliation{Beijing Computational Science Research Center, Beijing 100094, China}

\author{Masaki Sano}
\affiliation{Department of Physics, The University of Tokyo, Hongo 7-3-1, Tokyo 113-0033, Japan}

\date{\today}

\begin{abstract}
We study the collective dynamics of elongated swimmers in a very thin fluid layer by devising long, filamentous, non-tumbling bacteria. 
The strong confinement induces weak nematic alignment upon collision, which, for large enough density of cells, gives rise to global nematic order.
This homogeneous but fluctuating phase, observed on the largest experimentally-accessible scale of millimeters,
 exhibits the properties predicted by standard models for flocking such as the Vicsek-style model of polar particles 
with nematic alignment: true long-range nematic order and non-trivial giant number fluctuations.
\end{abstract}


\maketitle

Collective motion of self-propelled elements, as seen in bird flocks, fish schools, bacterial swarms, etc., is so ubiquitous that it 
has driven physicists to search for its possibly universal properties \cite{VICSEK-REVIEW,SR-REVIEW,MARCHETTI}. 
If generic and robust features of such active matter systems exist, they should also be present in the emergent 
phenomena observed in simple models. Evidence for such universality has been provided by
many theoretical and numerical studies of dry active matter systems where local alignment competes with noise, following the seminal works by Vicsek {\it et al.} \cite{VICSEK}, Toner, Tu , and Ramaswamy {\it et al.} \cite{TONER,TONER2,TONER3,TTR-REVIEW,RST2003}.
It was notably understood that the transition to orientational order/collective motion, in this context, is best described as a phase-separation
between a disordered gas and an ordered `liquid' separated by a coexistence phase whose nature depends on the symmetries of the
system \cite{MARCHETTI,GC-PRL,CHATE-PRE,NEMA-PRL,GINELLI,NGO,ST-PRL,SCT-PRL}.
The homogeneous but highly fluctuating liquid phase is characterized by unique properties often different from those of equilibrium orientationally-ordered phases. In particular, the crucial coupling between the order and the density fields generates anomalously-large number 
fluctuations from the algebraic correlations of orientation and density \cite{TONER,TONER2,TONER3,TTR-REVIEW,RST2003}.
 
Such `giant' number fluctuations (GNF), being relatively easy to measure experimentally, 
have become the landmark signature of orientationally-ordered active matter. 
Several experimental studies have indeed searched for GNF
using controllable systems simpler than bird flocks and fish schools such as biofilaments driven by molecular motors \cite{SCHALLER2}, colloids consuming electric energy \cite{BRICARD}, shaken granular materials \cite{NARAYAN,DESEIGNE,KUMAR}, monolayers of fibroblast cells \cite{DUCLOS}, and common bacteria \cite{ZHANG,WENSINK}. 
However, none of these experiments has been fully convincing in demonstrating the presence of {\it bona fide} GNF as predicted from the works of Toner, Tu, Ramaswamy {\it et al.} \cite{TTR-REVIEW,TONER3,MARCHETTI}, and observed in Vicsek-style models \cite{GC-PRL,CHATE-PRE,GINELLI,NGO}.
These GNF have to be discussed in a 
fluctuating phase with global long-range orientational order and are distinct
from the trivial, non-asymptotic ones present in the case of phase-separation into dense clusters sitting in a disordered sparse gas.
In some experiments, only normal number fluctuations were found 
\cite{BRICARD,SCHALLER2}. In others, GNF were reported for systems {\it not} in the fully ordered phase \cite{ZHANG,MYXO,NARAYAN,DESEIGNE,SCHALLER2, DUCLOS}. 
Finally, Refs.\cite{WENSINK,KUMAR} show some evidence of GNF only in numerical models of the experiments described. (Detailed interpretations on the above experiments are given in the Supplemental Material \cite{SUPP}.)
Difficulties and pitfalls indeed abound to observe unambiguous Toner-Tu-Ramaswamy phenomena:
very large systems are typically needed; external boundaries thus prevent their observation; it is often difficult to distinguish the coexistence phase from the liquid phase, leading one to
confuse the non-asymptotic fluctuations due to clustering with the GNF of the orientational liquid; strong steric interactions 
in dense systems may overcome alignment effects; additional long-range interactions may tame density fluctuations.

In this Rapid Communication, we report a biological system that constitutes an experimental realization of an orientationally-ordered active liquid 
as described by Toner, Tu, Ramaswamy {\it et al} \cite{TONER,TONER2,TONER3,TTR-REVIEW,RST2003}.
Specifically, we study the collective dynamics of long, filamentous, non-tumbling bacteria swimming in a very thin fluid layer between walls. 
Strong confinement and the high aspect ratio of cells induce weak nematic alignment upon collision, which gives rise to global nematic order at sufficiently high density of cells.
We show that this homogeneous but fluctuating ordered phase, observed on the largest experimentally-accessible scale of millimeters,
 exhibits the same properties as the Vicsek-style model of polar particles with nematic alignment \cite{GINELLI}: true long-range nematic order and non-trivial GNF.

The collective behavior of bacteria is a vast topic of research with obvious and crucial biological interest. 
Bacteria have also been widely used by physicists as attractive active matter systems.
Both crawling/sliding and swimming bacteria have been used, 
but so far, no very long-range ordering/collective motion has been observed. 
Sliding myxobacteria, for example, align, collide and form very dense ordered clusters, but these clusters are of limited size,
being easily destroyed upon collision and moving in directions \cite{MYXO}. 
{\it Bacillus subtilis} swimming/swarming on agar surfaces form loose ordered clusters with anomalous density fluctuations, again of limited size and moving in various directions \cite{ZHANG}.
Dense suspensions of swimming cells typically give rise to `bacterial turbulence'
\cite{SOKOLOV,ZHANG,SOKOLOV2,WENSINK,GACHELIN}, 
{\it i.e.} a chaotic regime with a dominant length scale of about 10--20 cell lengths. Two factors are often invoked to explain this situation: (i) Long-range hydrodynamic interactions are theoretically known to destabilize ordered states \cite{SUBRAMANIAN,SAINTILLAN,SAINTILLAN2,SAINTILLAN3,SAINTILLAN4}; (ii) the aspect ratio of cells is too small to lead to strong alignment upon collision \cite{WENSINK}.
To prevent these two pitfalls, we devised a system of filamentous cells of {\it Escherichia coli} bacteria ({\it E. coli}) \cite{TAKEUCHI,MAKI} confined 
between two solid walls with a small, micron-size gap. 
In addition to the information provided below, full details on our experimental setup can be found in the Supplemental Material \cite{SUPP}.

The filamentous bacteria were obtained by incubating usual {\it E. coli} cells under the influence of the antibiotic cephalexin 
(20 $\mu {\rm g}/{\rm ml}$), which allows cell growth but inhibits cell division. 
We used a non-tumbling chemotactic mutant strain RP4979, insuring persistent motion.
Cells were transformed to express yellow fluorescent protein. 
The filamentous cells have flagella all around their bodies at the same density as usual bacteria, 
and are still able to swim \cite{TAKEUCHI,MAKI}. 
Because the swimming speed gradually decreases with their length \cite{MAKI}, 
we used the bacteria with a moderate body length of $\sim19 \pm 5$ $\mu {\rm m}$ ($\pm$: standard deviation)
which have an aspect ratio of about 25 (see the Supplemental Material \cite{SUPP} for cell length distribution). 

\begin{figure}[t]
\includegraphics[width=\columnwidth]{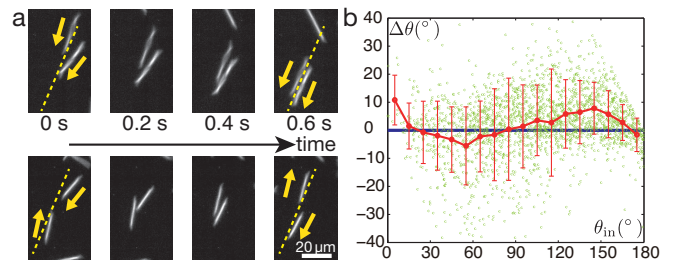}
\caption{
(a) Aligning collision events between two bacteria. Top: acute angle collision leading to alignment.
Bottom: obtuse angle collision leading to anti-alignment. Dashed line: mean outgoing angle. The mean incoming angle is not shown, as it is only slightly different from the outgoing angle. (b) Binary collision statistics: difference between the incoming angle and the outgoing angle $\Delta \theta$ vs the incoming angle $\theta_\mathrm{in}$. Green circles: 2214 individual collision events. Red circles: mean $\Delta\theta$ obtained via binning $\theta_\mathrm{in}$. Error bars: standard deviation. Collision events with $|\Delta\theta| >40^\circ$ are not shown for visibility. See the Supplemental Material for details \cite{SUPP}.
}
\label{fig1}
\end{figure}

The suspension of filamentous cells, after concentration, was sandwiched between a coverslip and a polydimethylsiloxane (PDMS) plate without any spacers to make as thin a chamber as possible, except for some wells as suspension reservoirs. We thus achieved a gap of about $\sim2$ $\mu {\rm m}$. Such a strong confinement contributed to suppress the destabilizing fluid flow due to no-slip boundary conditions on the walls.
The confinement also helped preventing bacterial circular motion near solid walls \cite{LAUGA}: The hydrodynamic interactions with each wall compensated each other \cite{SWIECICKI}, enabling our bacteria to swim straight over the longest distances (millimeters) considered below. 
Note that the gap width required for straight motion is larger for longer bacteria, so the use of the filamentous cells made it easier to design our experimental setup.

\begin{figure*}[t!]
\includegraphics[width=2\columnwidth]{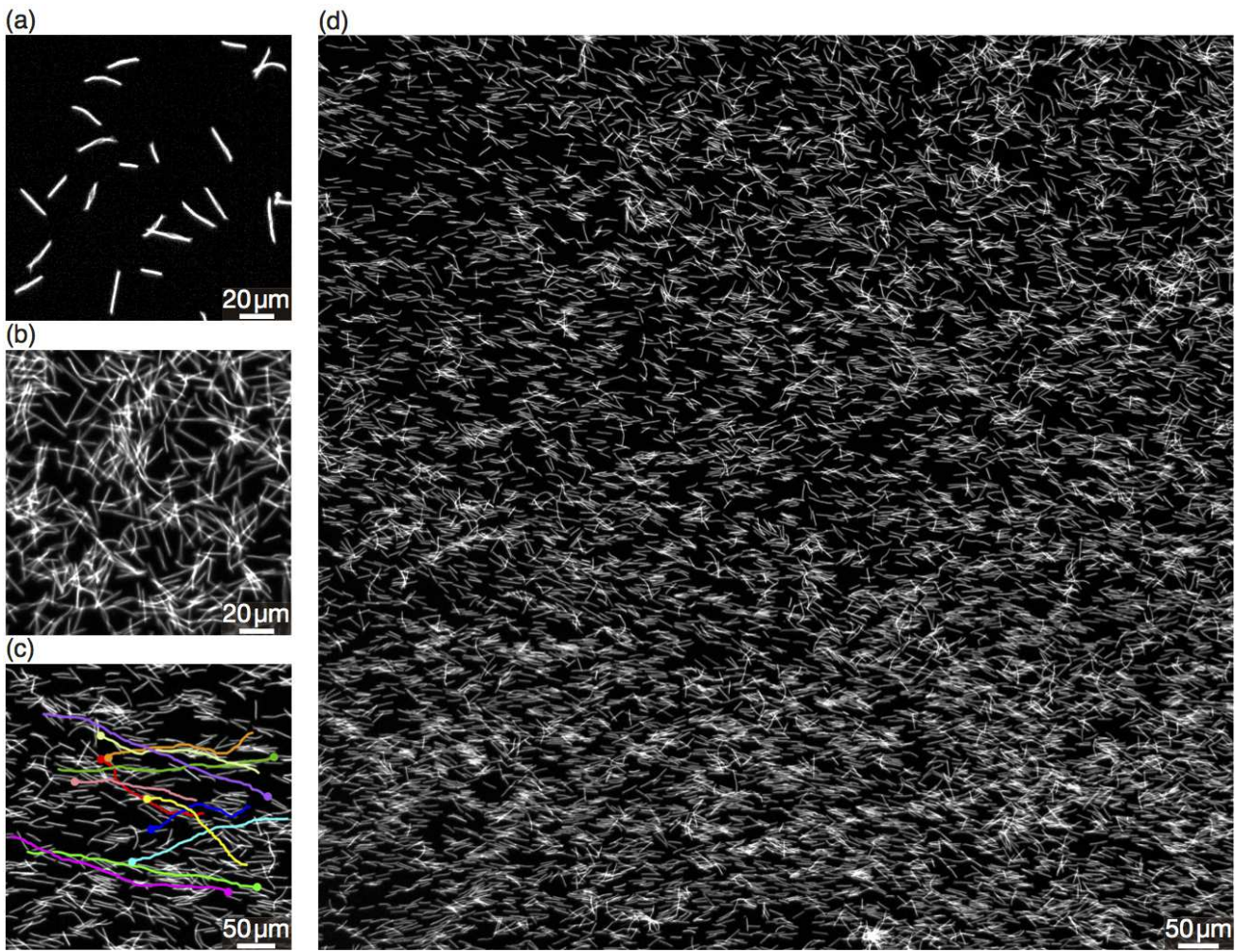}
\caption{
Typical snapshots.
(a) Zoom of the disordered phase at low density in a 2 $\mu{\rm m}$ thin experiment.
(b) Zoom of the disordered phase at high density in a 10 $\mu{\rm m}$ thick experiment.
(c) Zoom of the nematically-ordered phase at high density in a thin experiment with superimposed, manually tracked,
10-second trajectories of a few cells.
(d) Full field of view in the same experiment as in (c). See the Supplemental Material movies \cite{SUPP}.
[Note that the resolution of the figures are lowered for uploading this manuscript to arXiv.]
}
\label{fig2}
\end{figure*}

\begin{figure*}[t!]
\includegraphics[width=2\columnwidth]{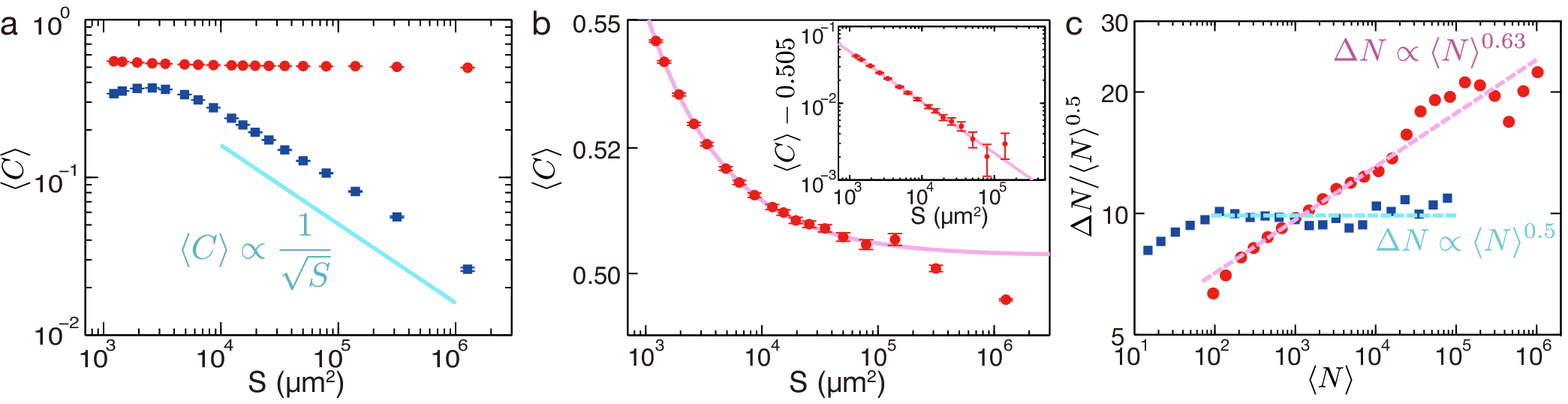}
\caption{
(a) Log-log plot of nematic order parameter $\langle C\rangle$ vs area $S$ of the ROI. Red circles: the globally nematically ordered state at high density in a very thin sample shown in Figs.~\ref{fig2}c,d. Blue squares: the disordered state at low density. Cyan solid line: slope of exponent $-0.5$ as a guide to the eye \cite{NOTE3}. The nematic order $\langle C\rangle$ stays at high values over the whole field of view in the ordered state (the red data).
(b) Same data as in (a) for the ordered state in a magnified range (log-log scale). The curvature in this log-log plot indicates {\it slower decay than a power law}. The last 3 points were excluded from the fit because they were not reliable due to longer correlation times and inhomogeneities at such large-scales. Inset: the same data from which the estimated asymptotic value of $C_\infty=0.505$ has been subtracted (log-log scale). Magenta solid lines: fit $\langle C \rangle=C_\infty +k S^\beta$ with $C_\infty=0.505$, $\beta=-0.66$, and $k=4.6$. Error bars in (a) and (b): standard error.
(c) Scaling of number fluctuations $\Delta N/\sqrt{\langle N\rangle}$ vs $\langle N\rangle$ on the log-log scale. Blue squares: normal fluctuations in the disordered, low density phase.
Red circles: anomalous, `giant' fluctuations recorded  in the high-density nematically-ordered state of Figs.~\ref{fig2}c,d. Cyan dashed line: normal fluctuations $\Delta N \propto \langle N \rangle^{0.5}$ as a guide to the eye. Magenta dashed line: fitted curve $\Delta N \propto \langle N \rangle^{0.63}$ for the ordered state.
}
\label{fig3}
\end{figure*}

After waiting for the initial fluid flow ---triggered when introducing the suspension--- to be suppressed, we captured movies by a CMOS camera (Baumer HXG40, $2048\times2048$ pixels, 12 bit) at 5 Hz through an inverted fluorescent microscope (Leica DMi8) with an objective lens (HC PL FLUOTAR, $10\times$, NA=0.30). The area of the field of view was $1.12\times1.12 \,{\rm mm}^2$,
a size limited by our will to be able to distinguish individual cells on the recorded images. The duration of the analyzed movies was 400 seconds (2000 frames), and there was no detectable change in bacterial lengths during the experiments (see the Supplemental Material \cite{SUPP}).
The microscope was equipped with an adaptive autofocusing system to reduce unwanted intensity fluctuations. We subtracted time-averaged dark current images from the obtained images and divided them by fluorescent images of homogeneous fluorophore (fluorescein) to calibrate the spatial inhomogeneity of the excitation light source. The dark current images were also subtracted from the fluorescent images beforehand (see the Supplemental Material \cite{SUPP}). Thanks to the permeability of PDMS to oxygen, typical experiments could be run for about 30 minutes without discernible
changes in the behavior of the cells. 

Our setup was thin enough to make it difficult for bacteria to cross each other without collisions. 
We have collected statistics on thousands of binary collisions using movies taken at a relatively low density of bacteria. 
(Details on the detection and analysis of collision events are given in the Supplemental Material \cite{SUPP}.).
Some clear events of `nematic alignment' upon collision are shown in Fig.\ref{fig1}a: Two bacteria incoming at some acute 
(obtuse) angle $\theta_\mathrm{in}$ end up parallel (antiparallel). Overall, however, alignment is weak, and many events do not result in such ideal nematic alignment with the outgoing angle $\theta_\mathrm{out} \simeq 0^\circ$ or $180^\circ$.
In Fig.\ref{fig1}b, we show that
the difference between incoming and outgoing angles $\Delta\theta=\theta_\mathrm{out}-\theta_\mathrm{in}$ is on average negative for $\theta_\mathrm{in}<90^\circ$ and positive for $\theta_\mathrm{out}>90^\circ$, characteristic of nematic alignment. We note also that our setup allows
for a significant fraction of events where bacteria cross each other undisturbed. (On the other hand, we recorded no events where alignment occurs {\it without} collision, ruling out hydrodynamic effects.) We believe this makes our system closer to Vicsek-style models 
where strong noise allow for non-alignment or even disalignment, something impossible in strictly two-dimensional experiments
\cite{ZHANG,WENSINK,MYXO,SOKOLOV}.

In experiments at low density of cells, or with a larger spacing ($\sim 10$ $\mu {\rm m}$) between the two surfaces, 
cells do not align enough to order on large scales (Figs.~\ref{fig2}a,b, and Supplemental Material \cite{SUPP}). But at high concentration
(average area fraction of $\sim0.25$), their collisions are so frequent that  
global nematic order emerges in spite of the weakness of alignment (see Ref. \cite{SUZUKI} for a similar situation).
This ordered phase is strongly fluctuating but statistically homogeneous, without clusters.
Bacteria then swim in opposite directions in approximately equal numbers (Figs.\ref{fig2}c,d, and Supplemental Material\cite{SUPP}). 
Since it is very difficult to determine the polarity $\theta$ of each bacterium at such large concentration,
a direct estimate of the nematic order parameter $Q=|\langle e^{2i\theta}\rangle|$ previously used even in experiments \cite{NISHIGUCHI} is out of reach, and we opted instead
for the `structure tensor' method used previously, e.g., for measuring 
the orientation of collagen fibers \cite{REZAKHANIHA}. Specifically, given an intensity-calibrated image $f(x,y)$, one calculates the following tensor over a given region of interest (ROI):
\begin{equation}
J = \left[ 
\begin{array}{cc} 
\langle \partial_x f , \partial_x f \rangle & \langle \partial_y f , \partial_x f \rangle \\
\langle \partial_x f , \partial_y f \rangle & \langle \partial_y f , \partial_y f \rangle 
\end{array}
\right]
\end{equation}
where $\langle g,h \rangle= \int\!\!\int_{\rm ROI} g\,h\,{\rm d}x \, {\rm d}y$. The eigenvalues $\lambda_{\rm min}$  and
$\lambda_{\rm max}$ of $J$ then give an estimate of the scalar nematic order parameter, called the `coherency parameter'
\begin{equation}
C \equiv \frac{ \lambda_{\rm max}-\lambda_{\rm min}}{\lambda_{\rm max}+\lambda_{\rm min}} \;,
\end{equation}
whereas the eigenvector corresponding to $\lambda_{\rm min}$ gives the orientation of the global nematic order in the ROI.

We have measured the nematic order parameter $\langle C \rangle$ for square ROIs of various area $S$, where the average is taken over both space and time. 
In the disordered phases observed either at low density or at high density but in a thicker layer of fluid, 
we find that $\langle C \rangle \sim 1/\sqrt{S}$, the same behavior as the conventional nematic order parameter $Q$ in the case of finite spatial correlation length (Fig.~\ref{fig3}a and \cite{NOTE3}).
In the ordered regime observed at the high density and with thin apparatus, on the other hand, we observe no topological defects and a very slow decay of the nematic order parameter.
As shown by the curvature of the log-log plot in Fig.~\ref{fig3}b, this decay is {\it slower than a power law}.
This is the signature of true long-range order.
As a matter of fact, an excellent fit of the data is an algebraic approach to some finite asymptotic value of $\langle C \rangle - C_\infty \sim S^\beta$,
with $C_\infty = 0.505$ and $\beta =-0.66$ (Fig.~\ref{fig3}b).
Similar finite-size scaling was found in the model studied in Ref. \cite{GINELLI}.

To quantify number fluctuations, instead of directly detecting each bacterium (again a difficult task), 
we binarized our images using the commonly-used Otsu's method \cite{OTSU}
and counted, in each square ROI centered at the field of view, the number of pixels $N(t)$ covered by bacteria at time $t$. 
The binarization process has the advantage of correcting for the slight differences in intensity resulting from variations of the height
of bacteria or fluctuations in the overall light intensity. 
On the other hand, it leads to small systematic underestimates in the case of overlapping cells.
We calculated the standard deviation $\Delta N = \sqrt{\langle [N(t)-\langle N \rangle]^2 \rangle}$ (all averages over time) for square ROI of various sizes.
In the disordered phase, we find normal fluctuations $\Delta N \sim {\langle N \rangle}^{0.5}$, but in the dense,
nematically-ordered phase, we estimate $\Delta N \sim \langle N \rangle^\alpha$ with $\alpha = 0.63(2) > 0.5$ \cite{NOTE2},
{\it i.e.} anomalous, giant fluctuations testifying to the presence of long-range correlations in the system (Fig.~\ref{fig3}c).

\begin{figure}[t!]
\includegraphics[width=\columnwidth]{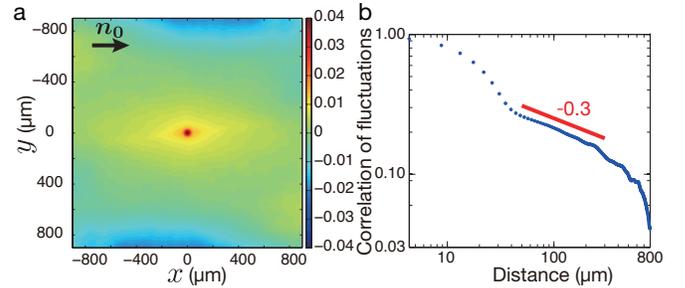}
\caption{
(a) Colormap of the correlation function $\mathrm{Corr}(\bm{R})$ of the director fluctuations $\bm{\delta n_\perp}=\bm{n}-\bm{n_0}$. The global mean director $\bm{n_0}$ is aligned in the $x$-direction.
(b) Log-log plot of the correlation function $\mathrm{Corr}(\bm{R})$ in the longitudinal direction (along $\bm{n_0}$). The red solid line is a slope with the exponent $-0.3$ just to guide the eye.
}
\label{fig4}
\end{figure}

We also measured correlations of fluctuations in the director $\bm{n}$ of the long-range nematic phase in our experiment.
From the structure tensor analysis, we have the local director field $\bm{n}$ and thus we can calculate the two-point correlation function \cite{LANDAU} of local director deviations from the global order $\bm{\delta n_\perp}=\bm{n}-\bm{n_0}$,
\begin{equation}
\mathrm{Corr}(\bm{R}) \equiv \langle \langle \delta n_\perp(t,\bm{r}) \delta n_\perp(t,\bm{r}+\bm{R}) \rangle_{\bm{r}} \rangle_t \;,
\end{equation}
where $\bm{n_0}$ is the global director obtained by spatially averaging $\bm{n}$ and $\delta n_\perp$ is a signed norm of $\bm{\delta n_\perp}$,
which is shown in Fig.~\ref{fig4}a (see the Supplemental Material \cite{SUPP} for details). In the longitudinal direction along $\bm{n_0}$, $\mathrm{Corr}(\bm{R})$ decays algebraically 
from the cell length up to the scale where the inhomogeneities of the setup are more pronounced or the constraint $\langle \delta n_\perp(t,\bm{r}) \rangle_{\bm{r}}=0$ makes $\mathrm{Corr}(\bm{R})$ become negative (Fig.~\ref{fig4}b).
These algebraic correlations in fluctuations can be associated with the Nambu-Goldstone mode and GNF \cite{TONER,TONER2,TONER3,TTR-REVIEW}.

A few comments are in order. Our system can be seen as a collection of self-propelled rods without velocity reversals that align nematically.
It should thus be compared {\it a priori} to the Vicsek-style model of polar particles with nematic interactions studied in Ref. \cite{GINELLI}. Indeed,
this model was shown to have true long-range nematic order over all numerically tested scales, as well as GNF with a scaling exponent of
$\alpha\simeq 0.75$. Our experimental findings are thus in full qualitative if not quantitative agreement with \cite{GINELLI}.
Our estimate of $\alpha$ is somewhat smaller, but this could be ascribed to both the limited range of accessible statistically-significant scales and/or excluded volume effects, which, after all, rule most if not all interactions in our system.

Our findings, like those of Ref. \cite{GINELLI}, challenge existing theoretical works. The linear
theory of Ramaswamy {\it et al.} for active nematic phases \cite{RST2003,TTR-REVIEW} predicts quasi long-range order in two dimensions 
({\it i.e.} an algebraic decay of nematic order to zero with increasing system size), 
and GNF with a scaling exponent of $\alpha=1$. At the nonlinear level, a perturbative renormalization group treatment has been performed
for active nematics without the density field, and concluded that the linear predictions should hold \cite{MSR2010}. But, as already pointed out in Refs. \cite{TTR-REVIEW,MSR2010}, nonlinear effects, especially some involving the density field, could change all this.

Regarding GNF, the value of $\alpha$ in the Toner-Tu-Ramaswamy orientationally-ordered phases is still the matter of debate, even in the case of polar flocks.
The value predicted by Toner and Tu in Refs. \cite{TONER,TONER2}, $\frac{4}{5}$, may 
not be exact, as originally claimed \cite{TONER3}, and it was only approximately confirmed numerically on the original Vicsek model in Refs. \cite{GC-PRL,CHATE-PRE}. For `pure' active nematics (apolar particles with fast velocity reversals), 
the latest numerical estimate of $\alpha$ is again around  $\frac{4}{5}$ \cite{NGO}, in contradiction with the linear theory. Here and in Ref. \cite{GINELLI} a slightly smaller value was again found.

In fact, a legitimate question, raised in past works \cite{MARCHETTI}, is whether self-propelled `rods' constitute an entirely different class
from polar flocks and active nematics. 
Their globally nematic phase can be seen as the superposition of two polar systems exchanging particles at some rate. As remarked in Ref. \cite{GINELLI}, this rate is low, and it defines a finite but large time/length, over which particles go in one of the two main directions defining the global nematic order. In our experiment, this length scale is certainly larger than our field of view, and thus, like in Ref. \cite{GINELLI}, we are unable to probe system sizes much larger than it.

Although the theoretical issues outlined above should still be resolved, our results provide the first unambiguous, large-scale, experimental evidence of the characteristic properties of order and fluctuations in globally-ordered homogeneous active phases predicted by the standard models of aligning 
self-propelled particles. In this context, future work will focus on obtaining better control on the density of bacteria so as to be able to study 
the transition to nematic order.
At the biological level, one could speculate that the long-range correlations put forward here might provide 
a means to collectively probe scales far beyond the individual cell's capacity. 

\ \linebreak
%\begin{acknowledgments}
We thank I. Kawagishi for providing the {\it E. coli} strain, Y. T. Maeda for transforming bacteria and reading the manuscript, K. A. Takeuchi for the use of the microscope, and K. Kawaguchi for discussion.
This work was supported by a Grant-in-Aid for the Japan Society for Promotion of Science (JSPS) Fellows (Grant No. 26-9915), the JSPS Core-to-Core Program ``Non-equilibrium dynamics of soft matter and information,''  KAKENHI (Grants No. 25103004, ``Fluctuation \& Structure'' and No. 26610112) from MEXT, Japan, and the French ANR project ``Bactterns''.
%\end{acknowledgments}

\bibliographystyle{apsrev4-1}

%\begin{widetext}
%\onecolumn
\clearpage
\onecolumngrid
%\appendix
\beginsupplement
\section*{Supplemental Material}
\subsection{Experimental procedure}
\begin{figure}[h]
\includegraphics[width=0.5\columnwidth]{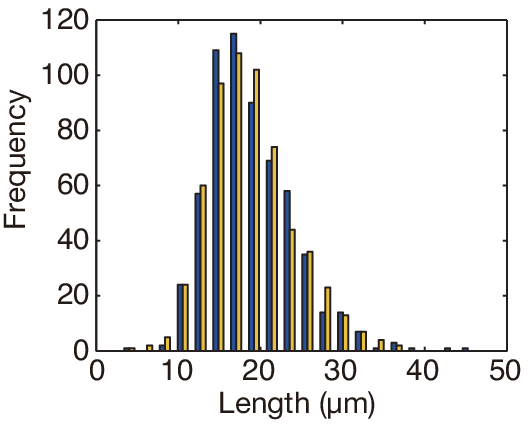}
\caption{Distribution of bacteria lengths before (blue) and after (yellow) one typical experiment. 
}
\label{SM_Fig_LengthDistribution}
\end{figure}

We used a non-tumbling chemotactic mutant strain of {\it Escherichia coli} (RP4979, $\mathrm{\Delta cheY}$, \cite{sSCHARF}) that had been transformed to express yellow fluorescent protein (plasmid: pZA3R-YFP). The strain RP4979, unlike a wild type stain RP437, lacks CheY protein responsible for flagella rotational switch. This CheY deleted mutant exclusively rotates flagella in counterclockwise (CCW) direction and swim persistently without tumbling. The bacteria taken from a frozen stock were grown overnight for 16 hours in Luria Broth (LB) with a selective antibiotic (chloramphenicol 33 $\mathrm{\mu g/ml}$) shaken at 200 rpm at 30 $\mathrm{^\circ C}$. Then this culture was diluted 100-fold in 10ml of Tryptone Broth (TB, 1 wt\% tryptone and 0.5 wt\% NaCl) with the selective antibiotic (chloramphenicol 33 $\mathrm{\mu g/ml}$) and incubated in a 125 ml flask shaken at 200 rpm at 30 $\mathrm{^\circ C}$. After 2 hours, the antibiotic cephalexin was added at the final concentration 20 $\mathrm{\mu g/ml}$ and we continued growing bacteria for another 3 hours to obtain filamentous cells. Although lengths of cells can be controlled by varying the duration of incubation after adding the antibiotic, their swimming speed gradually decreases with their lengths. We chose moderate body lengths $\sim19 \pm 5$ $\mu {\rm m}$ ($\pm$: standard deviation) to obtain cells with both sufficient nematic interactions and sufficient swimming speed (Fig.~\ref{SM_Fig_LengthDistribution}). To concentrate the obtained suspension, 1ml of the suspension was mildly passed through a membrane filter with 0.22 $\mathrm{\mu m}$ pores (Merck Millipore, Isomer GTBP01300) and we retrieved the concentrated suspension from the membrane.

This concentrated suspension was placed on a coverslip (MATSUNAMI, thickness 0.12-0.17 mm) and then sealed with a PDMS plate for observation. The PDMS plate was patterned with some wells which the excess fluid can escape into and can work as bacterial reservoirs. We used PDMS because it transmits oxygen required to sustain higher motilities of bacteria. Prior to putting the suspension onto the coverslip, the coverslip and the PDMS plate were soaked in 1 wt\% bovine serum albumin (BSA) solution for more than 1 hour in order to prevent the bacteria from sticking on the surfaces. To reduce the gap width, we slightly pressed the PDMS plate. After waiting for the initial fluid flow to be suppressed, we captured movies by a CMOS camera (Baumer HXG40, 2048×2048 pixels, 12 bit) at 5 Hz through an inverted fluorescent microscope (Leica DMi8 with Adaptive Focus Control) with an objective lens (Leica HC PL FLUOTAR, $10\times$, NA=0.30). The area of the field of view was $1.12\times1.12$ $\mathrm{mm^2}$. The microscope was equipped with an adaptive autofocusing system to reduce unfavorable intensity fluctuations. The duration of the analyzed movies was 400 seconds (2000 frames) and there was no detectable change of bacterial lengths during the experiments (Fig.~\ref{SM_Fig_LengthDistribution}). Hence this fact of no cell elongation together with inhibition of cell divisions insures that there is no gradual increase of the total number of bacteria or the area fraction, although in other usual bacterial experiments the number of bacteria doubles in approximately 20 minutes in the best conditions.

We note that the collision analysis shown in the Fig.1 in the main text were captured through another microscope (Nikon ECLIPSE TE2000-U) with different objective lenses (Nikon Plan Fluor ELWD, $40\times$, NA=0.60 for Fig.1a, and Nikon Plan Fluor, $10\times$, NA=0.30 for Fig.1b) .

%\newpage
\subsection{Image processing}
\begin{figure}[hp]
\includegraphics[width=\columnwidth]{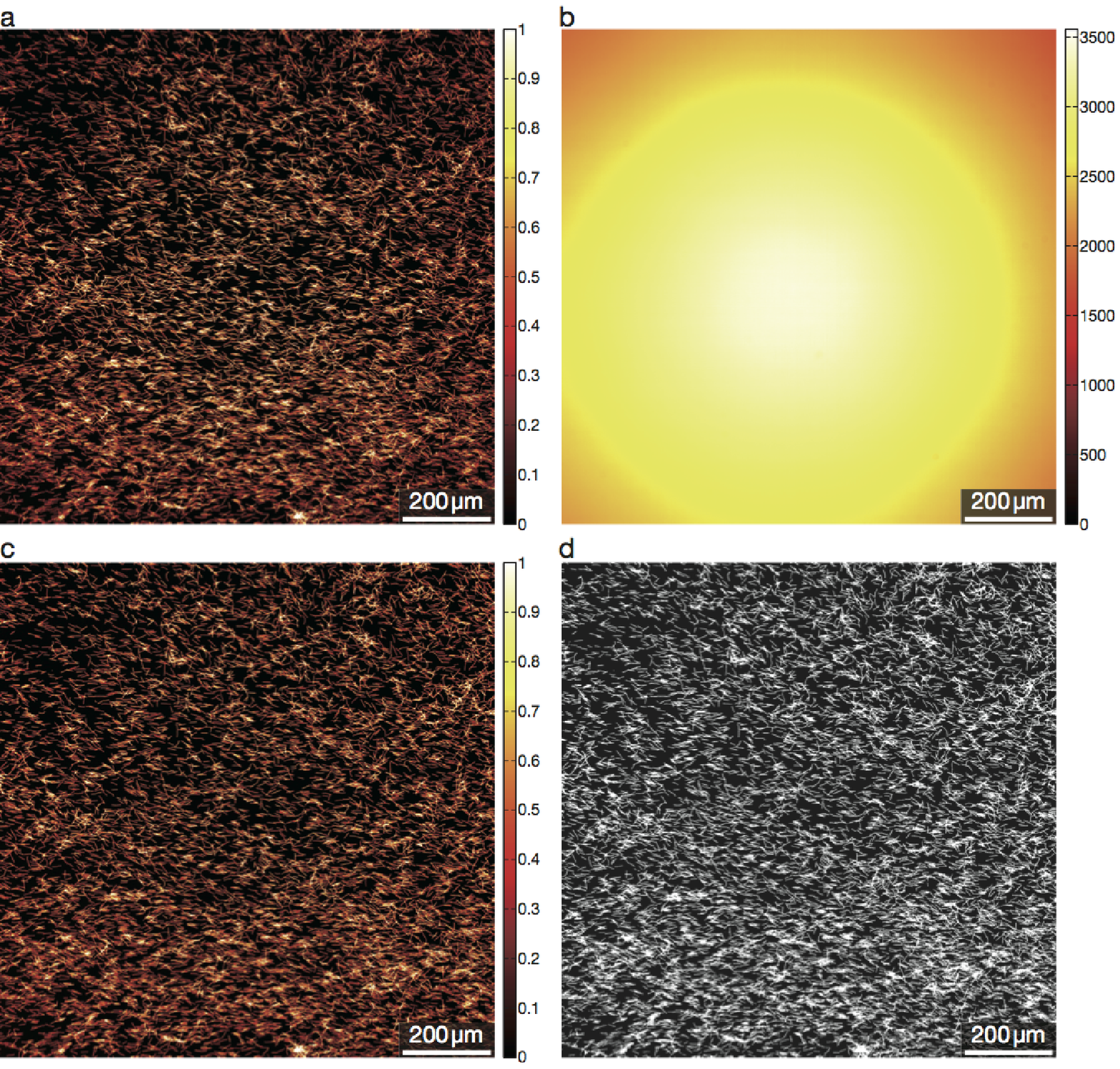}
\caption{(a) Raw image after the dark current subtraction. Intensity is adjusted in the range [0, 1] and contrast is enhanced for visibility. This image corresponds to the numerator in Eq.~(\ref{EqImageProcess}). (b) Intensity distribution of excitation light source for fluorescent microscopy. Fluorescence intensity of homogeneous fluorophore was obtained by the 12-bit camera and then after dark current subtraction it is used as the denominator of Eq.~(\ref{EqImageProcess}). (c) Calibrated image. Intensity is adjusted in the range [0, 1] and contrast is enhanced for visibility. (d) Binarized image used for analysis on number fluctuations.
[Note that the resolution of the figures are lowered for uploading this manuscript to arXiv.]
}
\label{SM_Fig_ImageProcess}
\end{figure}

Before estimating the coherency parameter, the number fluctuations, and the correlation functions, we first calibrated the intensity of the captured images as shown in Fig.~\ref{SM_Fig_ImageProcess}. We subtracted time-averaged dark current images from the obtained images of fluorescent {\it E.~coli} (Fig.~\ref{SM_Fig_ImageProcess}a) and divided them by fluorescent images of homogeneous fluorophore (fluorescein) (Fig.~\ref{SM_Fig_ImageProcess}b) to calibrate spatial inhomogeneity of the excitation light source. The dark current images were also subtracted from the fluorescent images beforehand. In summary,
\begin{equation}
\mbox{(calibrated image)} = \frac{ \mbox{(fluorescent {\it E.~coli} image)}-\mbox{(dark current image)}}{\mbox{(fluorophore image)}-\mbox{(dark current image)}} \;.
\label{EqImageProcess}
\end{equation}

Using these calibrated images (Fig.~\ref{SM_Fig_ImageProcess}c), we calculated the coherency parameter $C$, an estimate of the nematic order parameter, by using the structure tensor method \cite{sREZAKHANIHA}. We changed the box size $S$ and moved the boxes in such a way that each box does not overlap each other in order to ensure that all the pixels are used only once for the calculation, and then obtained mean values $\langle C \rangle$ by taking spatial and temporal averages.

When quantifying number fluctuations, we counted the number of pixels $N(t)$ covered by bacteria at time $t$ instead of directly detecting each bacterium. We binarized the calibrated images in order to correct the slight differences in intensity resulting from variations of the height of bacteria or fluctuations of the excitation light intensity (Fig.~\ref{SM_Fig_ImageProcess}d). We calculated a binarization threshold for each frame by applying commonly used Otsu's method \cite{sOTSU} in MATLAB and used their average value. The boxes used for calculating $N(t)$ were all centered at the field of view in order to avoid incorporating the spatial inhomogeneity of the setup into `number fluctuations'. In this analysis, a single filamentous bacterium corresponds to approximately $N\sim 100$ pixels.

\subsection{Collision analysis}
We investigated binary collisions and quantified interactions due to collisions. To decrease the number of multi-particle collisions, suspension of filamentous bacteria prepared in a way described above was diluted 3-fold in fresh Tryptone Broth with chloramphenicol. This diluted suspension was sandwiched between a coverslip and a PDMS plate in the same manner as in the experiment for the global nematic phase. We detected and tracked the center of mass of each bacterium from binarized images.  We defined the beginning of a collision as a merger event of two white objects in the binarized images that were isolated in the previous frame, and the end of the collision as a splitting event of that connected object. We inspected thousands of automatically detected merger events by eye and excluded multi-particle collisions. Thus we obtained 2204 collision events with accurate tracking.

\begin{figure}[tbhp]
\includegraphics[width=0.85\columnwidth]{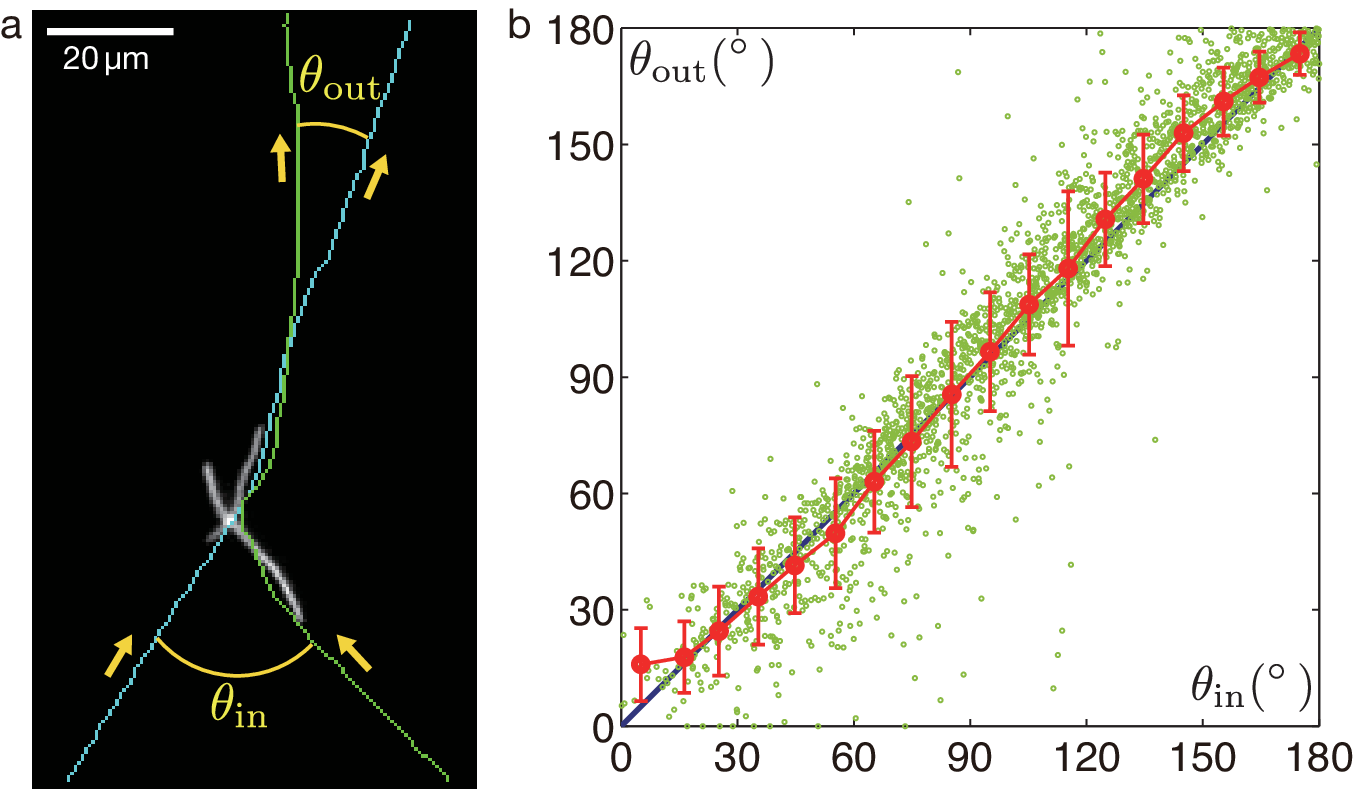}
\caption{(a) Snapshot of colliding bacteria. Trajectories of two bacteria are superimposed. The incoming angle $\theta_\mathrm{in}$ and the outgoing angle $\theta_\mathrm{out}$ of this collision is schematically shown. See Supplemental Movie 1 for this collision. (b) Outgoing angles $\theta_\mathrm{out}$ vs incoming angles $\theta_\mathrm{in}$. Green circles: individual collision events. All the analyzed 2204 collision events are shown here. Blue line: no interaction line ($\theta_\mathrm{out}=\theta_\mathrm{in}$). Red circles: mean $\theta_\mathrm{out}$ obtained via binning $\theta_\mathrm{in}$. Error bars: standard deviation.}
\label{SM_Fig_CollisionStatistics}
\end{figure}

Incoming angles $\theta_\mathrm{in}$ and outgoing angles $\theta_\mathrm{out}$ are calculated as angles formed by instantaneous velocity vectors of two bacteria just before and after the collision events respectively (Fig.~\ref{SM_Fig_CollisionStatistics}a). The velocity vectors are calculated from differences of positions in two successive frames separated by 0.2 seconds. Collision events with duration longer than 15 frames (3 seconds) are defined as complete polar alignment events ($\theta_\mathrm{out}=0^\circ$). All the analyzed data are shown in Fig.~\ref{SM_Fig_CollisionStatistics}b. Red data points in Fig.~\ref{SM_Fig_CollisionStatistics}b are mean values and standard deviations of $\theta_\mathrm{out}$ calculated via binning $\theta_\mathrm{in}$ at every $10^\circ$.

We note that the number of observed events in acute angles is biased to be smaller than that in obtuse angles, because durations of collisions with acute angles are usually longer and the probability of forming multi-particle collisions is higher.

\clearpage

%\afterpage{\clearpage}
\subsection{Correlation function}
The correlation function of director fluctuations $\mathrm{Corr}(\bm{R})$ was calculated from the director field obtained by the structure tensor method. We took ROIs of $64\times64$ pixels shifted by 8 pixels in either direction (87.5 \% overlap, $249\times249$ boxes in total). We calculated the local direction $\theta(t, \bm{r})$ of the director field for each $64\times64$-pixel ROI from the structure  tensor method \cite{sREZAKHANIHA}.
Using $\theta(t, \bm{r})$, we can calculate a correlation function,
\begin{equation}
\langle \delta n_\perp(t,\bm{r}) \delta n_\perp(t,\bm{r}+\bm{R}) \rangle_{\bm{r}} \;,
\end{equation}
which is used for nematic liquid crystals \cite{sLANDAU}. As is often the case, we assume the deviation $\bm{\delta n_\perp}=\bm{n}-\bm{n_0}$ of the local director $\bm{n}$ from the global mean director $\bm{n_0}$ is small. Because $|\bm{n}|^2=|\bm{n_0}|^2=1$ holds by definition,
\begin{eqnarray}
|\bm{n}|^2&=&|\bm{n_0}|^2+\bm{n_0}\cdot \bm{\delta n_\perp} + |\bm{\delta n_\perp}|^2\\
&\simeq& 1+ \bm{n_0}\cdot \bm{\delta n_\perp}\;,
\end{eqnarray}
and hence $\bm{n_0}\cdot \bm{\delta n_\perp}=0$ holds, meaning the fluctuations of the director field have only transverse components. The transverse component $\delta n_\perp$ of $\bm{\delta n_\perp}$ can be obtained as
\begin{equation}
\delta n_\perp (t,\bm{r})= \sin \left[ \theta(t,\bm{r})-\frac{1}{A}\int \theta(t,\bm{r})d^2 \bm{r} \right] \;,
\end{equation}
where the integral is over the whole field of view and $A$ is its area. Then we calculated $\langle \delta n_\perp(t,\bm{r}) \delta n_\perp(t,\bm{r}+\bm{R}) \rangle_{\bm{r}}$ and then averaged over time $t$,
\begin{equation}
\mathrm{Corr}(\bm{R})\equiv \langle \langle \delta n_\perp(t,\bm{r}) \delta n_\perp(t,\bm{r}+\bm{R}) \rangle_{\bm{r}} \rangle_t \;.
\end{equation}

To reduce the computational cost and to avoid calculating correlated successive images, we used every 20 frame of the movie (100 frames in total).

\subsection{Interpretations on existing experiments}
Here we detail existing experimental studies reporting `giant number fluctuations (GNF)', which is {\it not} actually measured in a {\it bona fide} Toner-Tu-Ramaswamy phase.
Let us state here again what we mean by `the {\it bona fide} GNF'. These GNF have to be discussed in a fluctuating phase with global long-range orientational order (the Toner-Tu-Ramaswamy phase) and are distinct from the trivial, non-asymptotic ones present in the case of phase-separation into dense clusters sitting in a disordered sparse gas. 

To the best of our knowledge, the only reported experiments that could satisfy all these conditions are the shaken granular particles of Deseigne {\it et al.} \cite{sDESEIGNE} and of Kumar {\it et al.} \cite{sKUMAR}. Unfortunately, these types of granular systems are composed of millimeter-sized particles and are limited by their boundaries, and long-range order cannot develop much. Moreover, Deseigne {\it et al.} could not bring their system to function deep in the ordered phase (as explained in Weber {\it et al.} \cite{sWEBER}) so that the fluctuations they recorded experimentally were those inherent to the transitional region. As a result, {\it bona fide} GNF could only be observed {\it in silico}, {\it i.e.} in the faithful numerical models presented in Kumar {\it et al.} \cite{sKUMAR} and in Weber {\it et al.} \cite{sWEBER}.

All other works we cite do not satisfy all conditions listed above.
\begin{itemize}
\item The shaken granular rods of Narayan {\it et al.} \cite{sNARAYAN} are again strongly affected by boundaries and suffer from the small system size, as is the case with the other granular experiments. Consequently, there exist many topological defects in the director field of the rods.
\item Zhang {\it et al.} \cite{sZHANG} are dealing with a globally disordered phase made of coexisting finite-size ordered clusters moving around in various directions. Even inside such ordered clusters, orientation/velocity correlations decay so rapidly at the length scale as small as $8\;\mathrm{\mu m} (\approx 2\;\mbox{bacterial body lengths})$ .
\item The myxobacteria of Peruani {\it et al.} \cite{sMYXO} show phase separation into dense, ordered clusters that do not order globally. Hence their GNF scaling is a trivial consequence of the phase-separation into dense clusters.
\item In the `mesoscale turbulence' paper of Wensink {\it et al.} \cite{sWENSINK}, GNF are only reported for a numerical model of rods, again in clustering/incoherent phases without long-range order.
\item In the fibroblast cells paper of Duclos {\it et al.} \cite{sDUCLOS}, GNF are measured in a domain phase with many topological defects and a finite correlation length. The correlation function of their orientation field were quite nicely fitted by an exponential function, which clearly demonstrates that there exists a finite correlation length and they only have short-range order. Furthermore, Duclos {\it et al.} observed GNF even at the lowest density with very small nematic order, which is 10 times more dilute than that of confluence. Therefore, it makes the dominant origin of these GNF rather obscure.
\item The actomyosin motility assay of the Bausch group \cite{sSCHALLER2} only shows GNF when the global flow is not straight. On the other hand, they report normal fluctuations when the system is most ordered at the beginning of the experiments, see the red data points in Fig.S5B of the Supplementary Material of Schaller {\it et al.} \cite{sSCHALLER2}.
\item The rolling colloids of Bricard {\it et al.} \cite{sBRICARD} only show normal fluctuations due to additional long-range hydrodynamic interactions.

\end{itemize}

\subsection{Movie descriptions}
\begin{description}
\item[Supplemental Movie 1] (1\_CollisionParallel.avi)\\
Binary collision leading to parallel alignment. Trajectories of two colliding bacteria are superimposed. The movie is played at the real speed.
\item[Supplemental Movie 2] (2\_CollisionAntiparallel.avi)\\
Binary collision leading to anti-parallel alignment. Trajectories of two colliding bacteria are superimposed. The movie is played at the real speed.
\item[Supplemental Movie 3] (3\_CollisionEffectiveTumbling.avi)\\
Binary collision with an obtuse incoming angle and an acute outgoing angle. Trajectories of two colliding bacteria are superimposed. This kind of rare events can be recognized as effective tumbling in self-propelled rods systems. The movie is played at the real speed.
\item[Supplemental Movie 4] (4\_disordered\_x3fast.mp4)\\
The disordered phase of filamentous cells of {\it E.~coli} at low density. The movie shows only the first 300 frames out of 2000 frames used for analysis. The movie is played at three times the real speed.
\item[Supplemental Movie 5] (5\_ordered\_x3fast\_OriginalSize.mp4)\\
The nematically-ordered phase of filamentous cells of {\it E.~coli} at high density. The movie shows only the first 300 frames out of 2000 frames used for analysis. The movie is played at three times the real speed.
\item[Supplemental Movie 6] (6\_ordered\_x3fast\_resized.mp4)\\
The nematically ordered phase of filamentous cells of {\it E.~coli} at high density. The same experiment as Supplemental Movie 5. The movie shows all the 2000 frames used for analysis, but the resolution is scaled down to $512\times512$ pixels from original $2048 \times 2048$ pixels just for reducing its data size. The movie is played at three times the real speed.
\item[Supplemental Movie 7] (7\_Tracking.avi)\\
Zoom of the nematically ordered phase at high density. The same experiment as Supplemental Movie 5 and 6. Manually tracked 10-second trajectories of 11 cells are overlaid on the captured movie. The movie is played at the real speed.
\end{description}

\bibliographystyle{apsrev4-1} 

\begin{thebibliography}{99}

\bibitem{VICSEK-REVIEW} T. Vicsek, and A. Zafeiris, Phys. Rep. {\bf 517}, 71 (2012).

\bibitem{SR-REVIEW} S. Ramaswamy, 
%The mechanics and statistics of active matter, 
Ann. Rev. Condens. Matter Phys. {\bf 1}, 323 (2010).

\bibitem{MARCHETTI} M.C. Marchetti, J.~F. Joanny, S. Ramaswamy, T.~B. Liverpool, J. Prost, M. Rao, R.~A. Simha, 
%Hydrodynamics of soft active matter. 
Rev. Mod. Phys. {\bf 85}, 1143 (2013).

\bibitem{VICSEK} T. Vicsek, A. Czir\'ok, E. Ben-Jacob, I. Cohen, and O. Shochet,
%Novel Type of Phase Transition in a System of Self-Driven Particles. 
Phys. Rev. Lett. {\bf 75}, 1226 (1995).

\bibitem{TONER} J. Toner, and Y. Tu,
%Long-range order in a two-dimensional dynamical XY model: How birds fly together. 
Phys. Rev. Lett. {\bf 75}, 4326 (1995).

\bibitem{TONER2} J. Toner, and Y. Tu, 
%Flocks, herds, and schools: A quantitative theory of flocking. 
Phys. Rev. E {\bf 58}, 4828 (1998).

\bibitem{TONER3} J. Toner, 
%Reanalysis of the hydrodynamic theory of fluid, polar-ordered flocks. 
Phys. Rev. E {\bf 86}, 031918 (2012). 

\bibitem{TTR-REVIEW} J. Toner, Y. Tu, and S. Ramaswamy. 
Annals of Physics (N.Y.) {\bf 318}, 170 (2005).

\bibitem{RST2003}S. Ramaswamy, R.~A. Simha, and J. Toner, Euro. Phys. Lett., {\bf 62}, 196 (2003). 
%Active nematics on a substrate: Giant number fluctuations and long-time tails

\bibitem{GC-PRL} G. Gr\'egoire, and H. Chat\'e, Phys. Rev. Lett. {\bf 92},  025702 (2004).

\bibitem{CHATE-PRE} H.~Chat\'e, F. Ginelli, G. Gr\'egoire, and F. Raynaud, Phys. Rev. E {\bf 77}, 046113 (2008).

\bibitem{NEMA-PRL} H. Chat\'e, F. Ginelli, and R. Montagne, 
%Simple model for active nematics: Quasi-long-range order and giant fluctuations, 
Phys. Rev. Lett. {\bf 96}, 180602 (2006).

\bibitem{GINELLI} F. Ginelli, F. Peruani, M. B\"ar, and H. Chat\'e, 
%Large-scale collective properties of self-propelled rods. 
Phys. Rev. Lett. {\bf 104}, 184502 (2010).

\bibitem{NGO} S. Ngo, A. Peshkov, I.~S. Aranson, E. Bertin, F. Ginelli, and H. Chat\'e, 
%Large-scale chaos and fluctuations in active nematics. 
Phys. Rev. Lett. {\bf 113}, 038302 (2014).

\bibitem{ST-PRL} A.P. Solon, and J. Tailleur, Phys. Rev. Lett. {\bf 111}, 078101 (2013).

\bibitem{SCT-PRL} A. P. Solon, H. Chat\'e, J. Tailleur, Phys. Rev. Lett. {\bf 114}, 068101 (2015).

\bibitem{SCHALLER2} V. Schaller, and  A.R. Bausch, 
%Topological defects and density fluctuations in collectively moving systems. 
Proc. Natl. Acad. Sci. U.S.A. {\bf 110}, 4488 (2013). Note that in this paper GNF were found when the system is 
disordered on large-scales, but not when it is ordered (cf. Figure S5 of the Supporting Information of this paper. See also the SupplementalMaterial \cite{SUPP}).

\bibitem{BRICARD} A. Bricard, J.-B. Caussin, N. Desreumaux, O. Dauchot, and D. Bartolo, 
%Emergence of macroscopic directed motion in populations of motile colloids. 
Nature {\bf 503}, 95 (2013).

\bibitem{NARAYAN} V. Narayan, S. Ramaswamy, and N. Menon, 
%Long-lived Giant Number Fluctuations in a Swarming Granular Nematic 
Science {\bf 317}, 105 (2007).

\bibitem{DESEIGNE} J. Deseigne, O. Dauchot, and H. Chat\'e, 
%Collective Motion of Vibrated Polar Disks. 
Phys. Rev. Lett. {\bf 105}, 098001 (2010). Note that the GNF found in this work were later found to be due to their measurement
being taken in the transitional region, not deep in the ordered phase. See C. A. Weber, {\it et al.}, Phys. Rev. Lett. {\bf 110}, 208001 (2013). See also the Supplemental Material \cite{SUPP}.

\bibitem{KUMAR} N. Kumar, H. Soni, S. Ramaswamy, and A.K. Sood, 
%Flocking at a distance in active granular matter. 
Nat. Commun. {\bf 5}, 4688 (2014).

\bibitem{DUCLOS} G. Duclos, S. Garcia, H.G. Yevick, and P. Silberzan,
%Perfect nematic order in confined monolayers of spindle-shaped cells
Soft Matter {\bf 10}, 2346 (2014). Note that the GNF found in this work were measured in a domain phase with many topological defects and with a finite correlation length. See also the Supplemental Material \cite{SUPP}.
%Note that the GNF in this work were found in short-ranged states with many oriented domains.

\bibitem{ZHANG} H.P. Zhang, A. Be'er, E.-L. Florin, and H.L. Swinney,
%Collective motion and density fluctuations in bacterial colonies. 
Proc. Natl. Acad. Sci. U.S.A. {\bf 107}, 13626 (2010).

\bibitem{WENSINK} H.H. Wensink, J. Dunkel, S. Heidenreich, K. Drescher, R.~E. Goldstein, H. L{\"o}wen, and J.~M. Yeomans, 
%Meso-scale turbulence in living fluids. 
Proc. Natl. Acad. Sci. U.S.A. {\bf 109}, 14308-14313 (2012).

\bibitem{MYXO} F. Peruani, J. Starruss, V. Jakovljevic, L. S{\o}gaard-Andersen, A. Deutsch, and M. B\"{a}r,
%Collective Motion and Nonequilibrium Cluster Formation in Colonies of Gliding Bacteria
Phys. Rev. Lett., {\bf 108}, 098102 (2012).

\bibitem{SUPP} See Supplemental Material at \url{http://link.aps.org/supplemental/10.1103/PhysRevE.95.020601} for experimental protocols, analysis methods, detailed interpretations on existing experimental works, and movies. 

\bibitem{SOKOLOV} A. Sokolov, I.S. Aranson, J.~O. Kessler, and R.E. Goldstein, 
%Concentration Dependence of the Collective Dynamics of Swimming Bacteria. 
Phys. Rev. Lett. {\bf 98}, 158102 (2007).

\bibitem{SOKOLOV2} A. Sokolov, and  I.S. Aranson, 
%Physical Properties of Collective Motion in Suspensions of Bacteria. 
Phys. Rev. Lett. {\bf 109}, 248109 (2012).

\bibitem{GACHELIN} J. Gachelin, A. Rousselet,  A. Lindner, and  E. Cl\'ement, 
%Collective motion in an active suspension of Escherichia coli bacteria. 
New J. Phys. {\bf 16}, 025003 (2014).

\bibitem{SUBRAMANIAN} G. Subramanian, and D.L. Koch,
%Critical bacterial concentration for the onset of collective swimming. 
J. Fluid Mech. {\bf 632}, 359 (2009).

\bibitem{SAINTILLAN} D. Saintillan, and M.J. Shelley,
%Emergence of coherent structures and large-scale flows in motile suspensions. 
J. R. Soc. Interface {\bf 9}, 571 (2012).

\bibitem{SAINTILLAN2} D. Saintillan, and M.J. Shelley,
Phys. Rev. Lett. {\bf 99}, 058102 (2007).

\bibitem{SAINTILLAN3} D. Saintillan, and M.J. Shelley,
Phys. Rev. Lett. {\bf 100}, 178103 (2008).

\bibitem{SAINTILLAN4} A. Lefauve, and D. Saintillan, 
Phys. Rev. E {\bf 89}, 021002 (2014).

\bibitem{TAKEUCHI} S. Takeuchi, W.R. Diluzio, D.B. Weibel, and G.M. Whitesides,
%Controlling the Shape of Filamentous Cells of Escherichia coli. 
Nano Lett. {\bf 5}, 1819 (2005).

\bibitem{MAKI} N. Maki, J.~E. Gestwicki, E.~M. Lake, L.~L. Kiessling, and J. Adler, 
%Motility and Chemotaxis of Filamentous Cells of Escherichia coli 
J. Bacteriol. {\bf 182}, 4337 (2000).

\bibitem{LAUGA} E. Lauga, W.R. DiLuzio, G.M. Whitesides, and H.A. Stone, 
%Swimming in circles: motion of bacteria near solid boundaries. 
Biophys. J. {\bf 90}, 400 (2006).
	
\bibitem{SWIECICKI} J.-M. Swiecicki, O. Sliusarenko, and D.B. Weibel,
%From swimming to swarming: Escherichia coli cell motility in two-dimensions. 
Integr. Biol. (Camb). {\bf 5}, 1490 (2013).

\bibitem{NISHIGUCHI} D. Nishiguchi, and M. Sano, 
%Mesoscopic turbulence and local order in Janus particles self-propelling under an ac electric field. 
Phys. Rev. E {\bf 92}, 052309 (2015).

\bibitem{REZAKHANIHA} R. Rezakhaniha, A. Agianniotis, J.~T.~C. Schrauwen, A. Griffa, D. Sage, C.~V.~C. Bouten, F.~N. van de Vosse, M.~Unser, and N. Stergiopulos, 
%Experimental investigation of collagen waviness and orientation in the arterial adventitia using confocal laser scanning microscopy. 
Biomech. Model. Mechanobiol. {\bf 11}, 461 (2012).

\bibitem{NOTE3} The data of the disordered state deviate from $\langle C \rangle \propto 1/\sqrt{S}$ in small $S$ because the number of ROIs without any bacteria ($C\simeq 0$) increases. 

\bibitem{SUZUKI} R. Suzuki, C.A. Weber, E. Frey, and A.R. Bausch, Nat. Phys. {\bf 11}, 839 (2015).

\bibitem{OTSU} N. Otsu. 
%A threshold selection method from gray-level histograms. 
IEEE Trans. Sys., Man., Cyber. {\bf 9}, 62 (1979).

\bibitem{NOTE2} Here the uncertainty means 95~\% confidence level. 

\bibitem{LANDAU} L.~D. Landau and E.~M. Lifshitz, {\it Statistical Physics}, 3rd Edition Part 1. (Elsevier, 1980) 

\bibitem{MSR2010} S. Mishra, R.~A. Simha, and S. Ramaswamy, J. Stat. Mech., P02003 (2010).
%A dynamical renormalization group study of active nematics

\end{thebibliography}

\begin{thebibliography}{99}
\bibitem{SCHARF} B.~E. Scharf, K.~A. Fahrner, L. Turner, and H.~C. Berg,
%Control of direction of flagellar rotation in bacterial chemotaxis,
Proc. Natl. Acad. Sci. USA, {\bf 9}, 201 (1998).

\bibitem{sREZAKHANIHA} R. Rezakhaniha, A. Agianniotis, J.~T.~C. Schrauwen, A. Griffa, D. Sage, C.~V.~C. Bouten, F.~N. van de Vosse, M.~Unser, and N. Stergiopulos, 
%Experimental investigation of collagen waviness and orientation in the arterial adventitia using confocal laser scanning microscopy. 
Biomech. Model. Mechanobiol. {\bf 11}, 461 (2012).
\bibitem{sOTSU} N. Otsu. 
%A threshold selection method from gray-level histograms. 
IEEE Trans. Sys., Man., Cyber. {\bf 9}, 62 (1979).

\bibitem{sLANDAU} L.~D. Landau and E.~M. Lifshitz, {\it Statistical Physics}, 3rd Edition Part 1. (Elsevier, 1980) 

\bibitem{sDESEIGNE} J. Deseigne, O. Dauchot, and H. Chat\'e, 
%Collective Motion of Vibrated Polar Disks. 
Phys. Rev. Lett. {\bf 105}, 098001 (2010). 

\bibitem{sKUMAR} N. Kumar, H. Soni, S. Ramaswamy, and A.K. Sood, 
%Flocking at a distance in active granular matter. 
Nat. Commun. {\bf 5}, 4688 (2014).

\bibitem{sWEBER} C.~A. Weber, T. Hanke, J. Deseigne, S. L\'eonard, O. Dauchot, E. Frey, H. Chat\'e,
%Long-Range Ordering of Vibrated Polar Disks
Phys. Rev. Lett. {\bf 110}, 208001 (2013).

\bibitem{sNARAYAN} V. Narayan, S. Ramaswamy, and N. Menon, 
%Long-lived Giant Number Fluctuations in a Swarming Granular Nematic 
Science {\bf 317}, 105 (2007).

\bibitem{sZHANG} H.P. Zhang, A. Be'er, E.-L. Florin, and H.L. Swinney,
%Collective motion and density fluctuations in bacterial colonies. 
Proc. Natl. Acad. Sci. U.S.A. {\bf 107}, 13626 (2010).

\bibitem{sMYXO} F. Peruani, J. Starruss, V. Jakovljevic, L. S{\o}gaard-Andersen, A. Deutsch, and M. B\"{a}r,
%Collective Motion and Nonequilibrium Cluster Formation in Colonies of Gliding Bacteria
Phys. Rev. Lett., {\bf 108}, 098102 (2012).

\bibitem{sWENSINK} H.H. Wensink, J. Dunkel, S. Heidenreich, K. Drescher, R.~E. Goldstein, H. L{\"o}wen, and J.~M. Yeomans, 
%Meso-scale turbulence in living fluids. 
Proc. Natl. Acad. Sci. U.S.A. {\bf 109}, 14308-14313 (2012).

\bibitem{sDUCLOS} G. Duclos, S. Garcia, H.G. Yevick, and P. Silberzan,
%Perfect nematic order in confined monolayers of spindle-shaped cells
Soft Matter {\bf 10}, 2346 (2014).

\bibitem{sSCHALLER2} V. Schaller, and  A.R. Bausch, 
%Topological defects and density fluctuations in collectively moving systems. 
Proc. Natl. Acad. Sci. U.S.A. {\bf 110}, 4488 (2013).

\bibitem{sBRICARD} A. Bricard, J.-B. Caussin, N. Desreumaux, O. Dauchot, and D. Bartolo, 
%Emergence of macroscopic directed motion in populations of motile colloids. 
Nature {\bf 503}, 95 (2013).


\end{thebibliography}

%\end{widetext}
\end{document}